\documentclass[journal]{IEEEtran}
\usepackage{booktabs}
\usepackage{xcolor}

\usepackage{cite}

\ifCLASSINFOpdf
   \usepackage[pdftex]{graphicx}
\else
\fi
\begin{document}
%
\title{A novel wireless sensor network topology with fewer links}
%
%
%

\author{Jie~Ding,
        Min-Yi~Wang, Qiao Wang,
        and Xin-Shan~Zhu
            \thanks{J. Ding and M. Y. Wang are with the School of Information Engineering,
                    Yangzhou University, Yangzhou 225127, China (e-mail: jieding@yzu.edu.cn, h07000113@126.com)}
            \thanks{Q. Wang is with the School of Information Science and Engineering, Southeast University,
                    Nanjing 210096, China (e-mail: qiaowang@seu.edu.cn)}%
            \thanks{X. S. Zhu is with the School of Electrical Engineering and Automation,
                    Tianjin University,  Tianjin 300072, China (e-mail: xszhu126@126.com)}
        }

%
%

\markboth{Journal of \LaTeX\ Class Files,~Vol.~3, No.~2, March~2014}%
{Ding \MakeLowercase{\textit{et al.}}: A novel wireless sensor network topology with fewer links}
%



\maketitle

\begin{abstract}
This paper, based on $k$-NN graph, presents symmetric $(k,j)$-NN graph $(1 \leq j < k)$,
a brand new  topology which could be adopted by a series of network-based structures.
We show that the $k$ nearest neighbors of a node exert disparate influence on guaranteeing network connectivity, and connections with the farthest $j$ ones among these $k$ neighbors are competent to build up a connected network, contrast to the current popular strategy of connecting all these $k$ neighbors. In particular, for a network with node amount $n$ up to $10^3$, as experiments demonstrate,
connecting with the farthest three, rather than all, of the five nearest neighbor nodes, i.e. $(k,j)=(5,3)$, can guarantee the network connectivity  in high probabilities. We further  reveal that more than $0.75n$ links or edges in $5$-NN graph are not necessary for the connectivity. Moreover, a composite topology combining symmetric $(k,j)$-NN and random geometric graph (RGG) is constructed for constrained transmission radii in wireless sensor networks (WSNs) application.
\end{abstract}

\begin{IEEEkeywords}
Network connectivity, Random geometric graph, Small-scale topology, Symmetric $(k,j)$-NN graph, Wireless sensor networks
\end{IEEEkeywords}

%
\IEEEpeerreviewmaketitle

\section{Introduction}
%
%
%
%
\IEEEPARstart{S}{ince} $1970$s, numerous technical literatures, given the demand of investigating wireless networks, concentrate on the problem of node proximity to its neighbors, or node degree. Assuming all nodes possess the same transmitting power in slotted ALOHA protocol, $6$ --- the average node degree --- is the ``magic number'' for maximizing the transmission distance of a single hop \cite{kleinrock1978optimum}. Takagi \emph{et al.} \cite{takagi1984optimal} reset the degree limit to $5$ or $7$. Considering adjustable transmission radius of individual node, the boundary is $6$ or $8$ in \cite{hou1986transmission}. For maximizing transmitting efficiency, Hajek \emph{et al.} \cite{hajek1983adaptive} indicated that each node should hold its transmission coverage for $3$ nearest neighbors in average.

\par For a large-scale network consists of $n$ nodes that uniformly distributed, it was pointed out by Xue \emph{et al.} in \cite{xue2004number} that the  network tends to be unconnected with probability one as $n$ goes to infinity, if the average node degree is less than $0.074\log n$. On the other hand, if the average node degree exceeds $5.1774\log n$, the  network tends to be connected  with  probability one. Later, Balister \emph{et al.} \cite{balister2005connectivity} improved the lower and upper limit to be $0.3043\log n$ and $0.5139\log n$ respectively. For small-scale network topology, Ni \emph{et al.} \cite{ni1994connectivity} concluded that connectivity would be guaranteed in high probability if a node possesses $6$ to $8$ neighbors.

\par It is worth mentioning that larger node degree will lead to more energy consumption caused by maintaining more number of links, as well as may increase the message collision probability if contention-based MAC protocols are adopted.
 Thus, it is very important to decrease the average node degree while safeguarding network connectivity. Based on $k$-NN graph, we build symmetric $(k,j)$-NN graph $(1 \leq j < k)$, a small-scale ($10^2$ to $10^3$ order of magnitude) network topology. Our researches demonstrate that the first $k$ nearest neighbors of a node exerts disparate influence on guaranteeing network connectivity. In particular, as simulations suggest, connections with
the farthest  three of the first five nearest nodes $(k, j)=(5,3)$ can competently build up a connected network. In symmetric $(5,3)$-NN graph, the average node degree decreases to around $4.4902$, which is significantly less than $6$ or $8$ in \cite{ni1994connectivity}. Given constrained transmission radius in WSNs applications, we further merge symmetric $(k,j)$-NN graph with random geometric graph (RGG)
without loss of connectivity. The composite topology boasts an amazing average node degree of $4.4316$. Its link gain, i.e. the number of saved links, reaches $0.78735n$, where $n$ is node amount.

\par The rest of the article is organized as follows. In Section~2, notions on symmetric $(k,j)$-NN graph as well as the combination of symmetric $(k,j)$-NN and RGG are briefly described. Section~3 demonstrates some numerical results and statistical analysis. Section~4 concludes the paper.

\section{Topology Description}

This section presents the mathematical model of our topology, i.e. $(k,j)$-NN graph, which is based on the common known $k$-NN graph. Consider constrained transmission radii in realistic applications, we propose give a composite topology which
combines a random geometry graph model.

\subsection{Symmetric $(k,j)$-NN Graph}

\begin{figure}[htbp]
  \centering
  \includegraphics[scale=0.4]{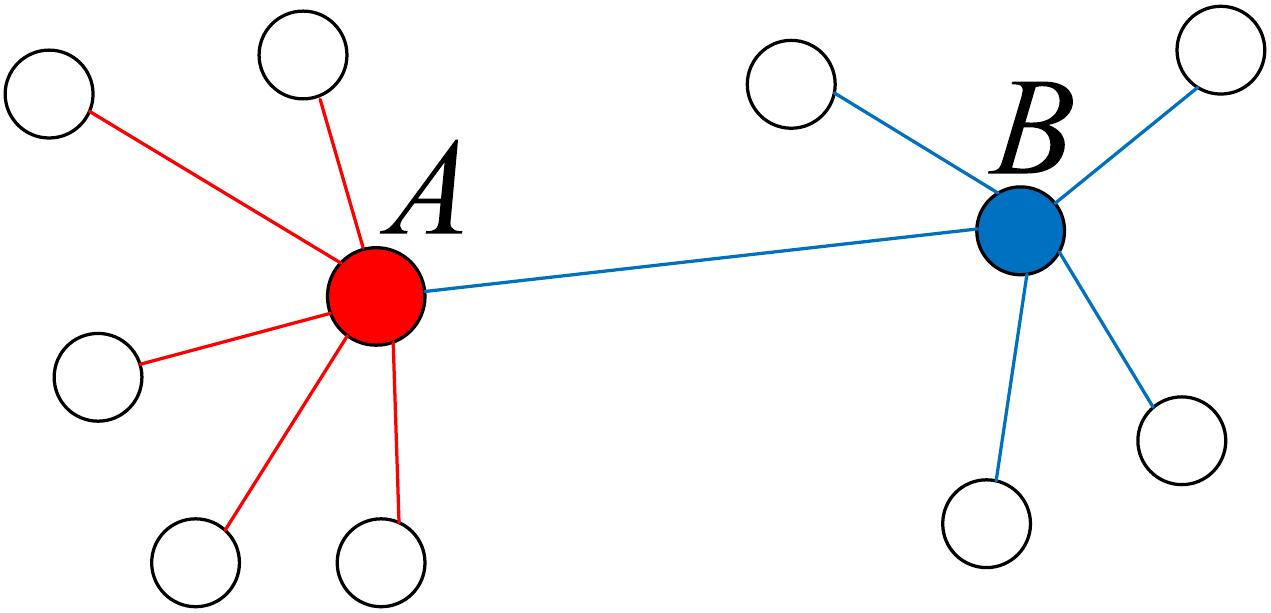}
  \caption{Symmetric $k$-NN graph ($k = 5$, but node $A$ has $6$ links)}\label{kNN}
\end{figure}

A $k$-NN graph in which each node connects its $k$ nearest neighbors, is a basic mathematical model for WSNs, where the sensors and links are modeled as nodes and edges. In this graph, a node, namely $A$, may be one of another node $B$'s $k$ nearest neighbors, but node $B$ is not in the first $k$ neighbor list of node $A$. Considering WSNs scenario that any two nodes could communicate with each other, we connect node $A$ and $B$. So a symmetric $k$-NN graph is created. Note that a node in symmetric $k$-NN graph may have more than $k$ connected neighbors (see Figure \ref{kNN}).

\par
\begin{figure}[htbp]
  \centering
  \includegraphics[scale=0.35]{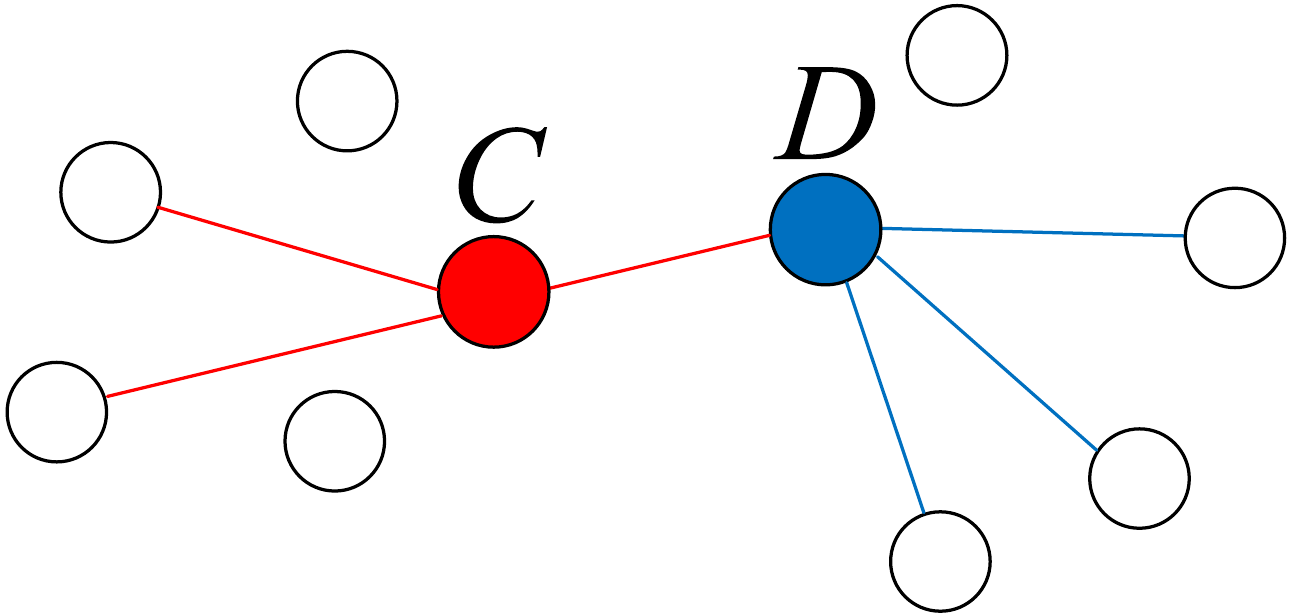}
  \caption{Symmetric $(k,j)$-NN graph ($k = 5$, $j = 3$, but node $D$ has 4 links)}\label{kjNN}
\end{figure}

 A $(k,j)$-NN graph $(1 \leq j < k)$ is a subgraph obtained by disconnecting each node and its $j-1$ nearest neighbors, i.e. removing the shortest $j-1$ edges or links for each node, in a symmetric $k$-NN graph. We consider the symmetric version of a $(k,j)$-NN graph. That is, in a symmetric $(k,j)$-NN graph, if a node (namely $C$) is in the first $j-1$ neighbor list of another node (namely $D$) while node $D$ does not belong to that of node $C$, we do not remove the edge between node $C$ and $D$, as shown in Figure \ref{kjNN}.
 In WSNs, the removal of a link can be simply carried out by idling this link. Clearly, the symmetric $(k,j)$-NN graph is still a subgraph of symmetric $k$-NN graph but has fewer edges or links. Next section will demonstrate that the subgraph enjoys almost the same connectivity probability. That is to say, the shortest $j-1$ links are redundant to connectivity and this is the reason why we remove them.

\subsection{The Combination of Symmetric $(k,j)$-NN Graph and RGG}
\par In the realistic applications of WSNs, the transmission radius of a node is constrained, which possibly results in failing to connect far distant neighbor nodes. To deal with this problem, we propose a composite topology which is based on the combination of symmetric $(k,j)$-NN graph and RGG.  We assume that each node have the same constrained transmission radius $r$ and they are uniformly distributed on a unit area square. For each node, if the population of its neighbors located in the disk centered at this node with radius $r$ is less than $k$, we connect this node and all these neighbors. Otherwise, if the population is more than or equal to $k$, then  the rule of symmetric $(k,j)$-NN strategy applies. For convenience, this composite topology is  denoted by $(k,j)$-NN-RGG.

\section{Numerical Results}
\par We assume that  all nodes are uniformly distributed in a unit-area square. Our experiments concentrate on small-scale ($10^2$ to $10^3$ order of magnitude) network topology. For each scenario, we repeatedly run a simulation $100$ times and record the mean value of corresponding variables.

\par
\begin{figure}[htbp]
  \centering
  \includegraphics[scale=0.6]{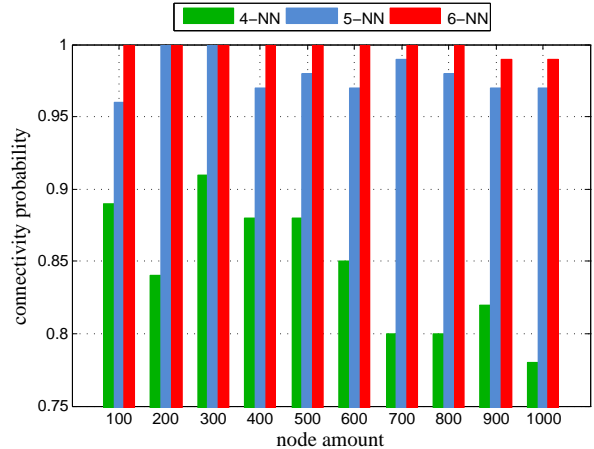}\\
  \caption{Connectivity probability: symmetric $4$-NN, $5$-NN, $6$-NN graph}\label{fig:c-prob}
\end{figure}

 First, we explore a feasible $k$ value for the symmetric $k$-NN graph model. A moderate $k$ is a balanced tradeoff between safeguarding network connectivity and cutting link overhead. We respectively record the connectivity probabilities of  networks while varying node amount from $100$ to $1000$. As Figure \ref{fig:c-prob} suggests, $k = 4$ deteriorates the connectivity probabilities. However, when $k$ is $5$, the network connectivity could be safeguarded in high probabilities. And $k = 6$ generates a perfect connected network in most cases, while $k \geq 7$ leads to excessive link overhead. Therefore, $k = 5$ and $k = 6$ are the candidates that we pick for further analysis.

\par
\begin{figure}[htbp]
  \centering
  \includegraphics[scale=0.6]{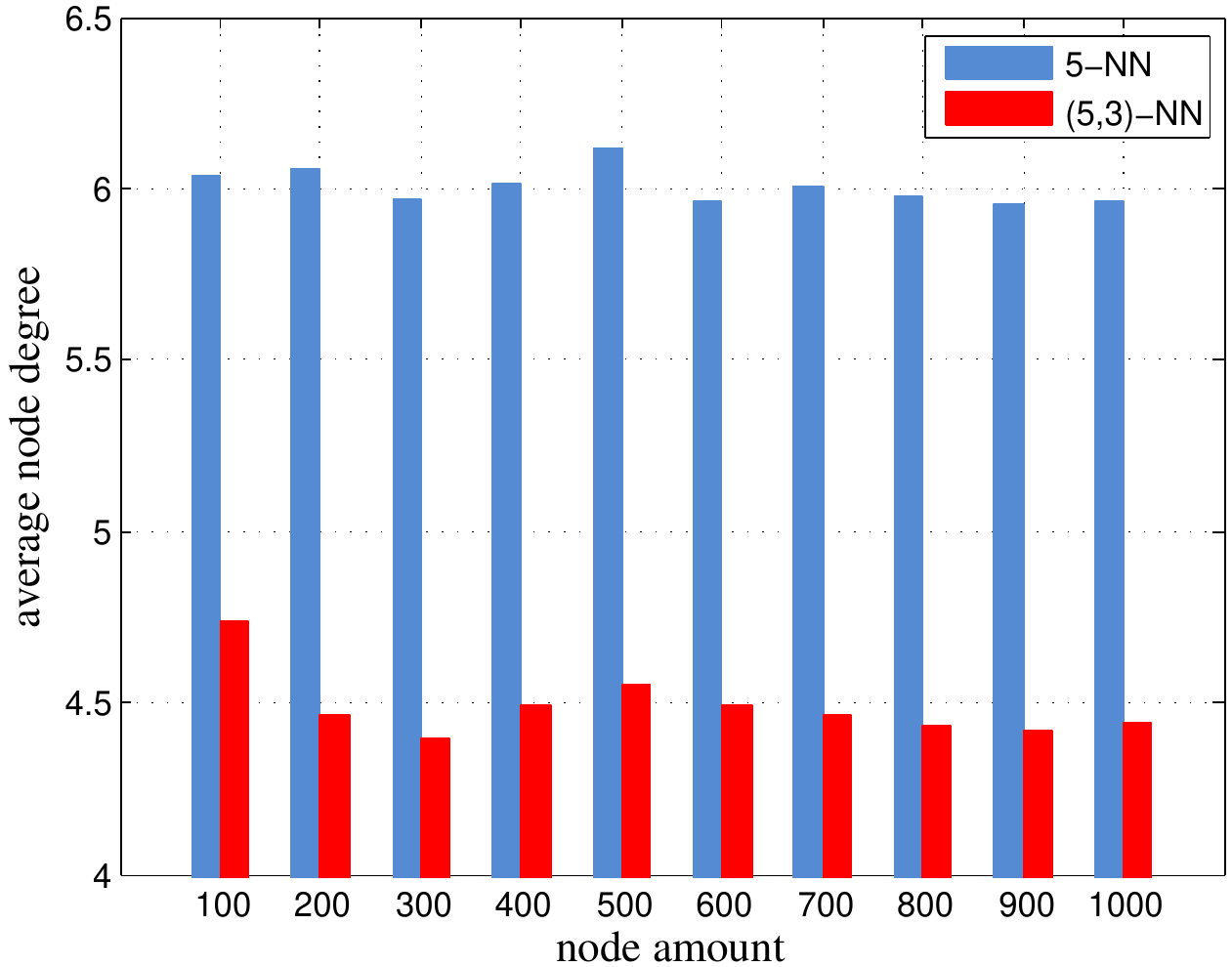}\\
  \caption{Average node degree: symmetric $5$-NN, $(5,3)$-NN graph}\label{fig:deg-51-53}
\end{figure}

 In the following, let us choose a suitable $j$ value. It is worth keeping in mind that larger node degree results in fatter chance of message collision. We aim to reduce  the  average node degree while guaranteeing network connectivity. Figure~\ref{fig:deg-51-53} shows that when $k$ is $5$, symmetric $(5, 3)$-NN graphs achieve around $4.4902$ average node degree --- the number rises to around $6.0063$ in symmetric $5$-NN graphs\footnote{A node in symmetric $k$-NN graph may have more than $k$ connected neighbors.} --- while guaranteeing network connectivity in high probability.\footnote{For node amount from $100$ to $1000$, the connectivity probabilities of symmetric $(5, 3)$-NN graph are all larger than $95\%$.} The link gain that a symmetric $(5, 3)$-NN graph holds reaches around $(6.0063-4.4902) \times \frac{n}{2} = 0.75805n$, where $n$ is node amount. Note that symmetric $(5, 2)$-NN graphs waste links, and symmetric $(5, 4)$-NN graphs fail to build up a connected network due to the lack of links.

\par A similar analysis could be made when $k$ is $6$. The symmetric $(6, 5)$-NN graph achieves amazing $3.3803$ average node degree compared to $7.1277$ in a symmetric $6$-NN graph, and thus the  link gain is around $(7.1277-3.3803) \times \frac{n}{2} = 1.8737n$. However, symmetric $(6, 4)$-NN graphs squander link resources and symmetric $(6, 6)$-NN graphs fail to span a connected network.

\begin{figure}[htbp]
  \centering
  \includegraphics[scale=0.6]{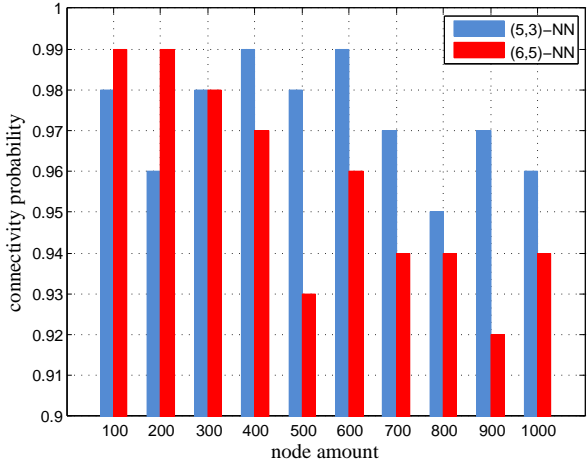}\\
  \caption{Connectivity probability: symmetric $(5,3)$-NN, $(6,5)$-NN graph}\label{fig:prob-53-65}
\end{figure}

\begin{figure}[htbp]
  \centering
  \includegraphics[scale=0.6]{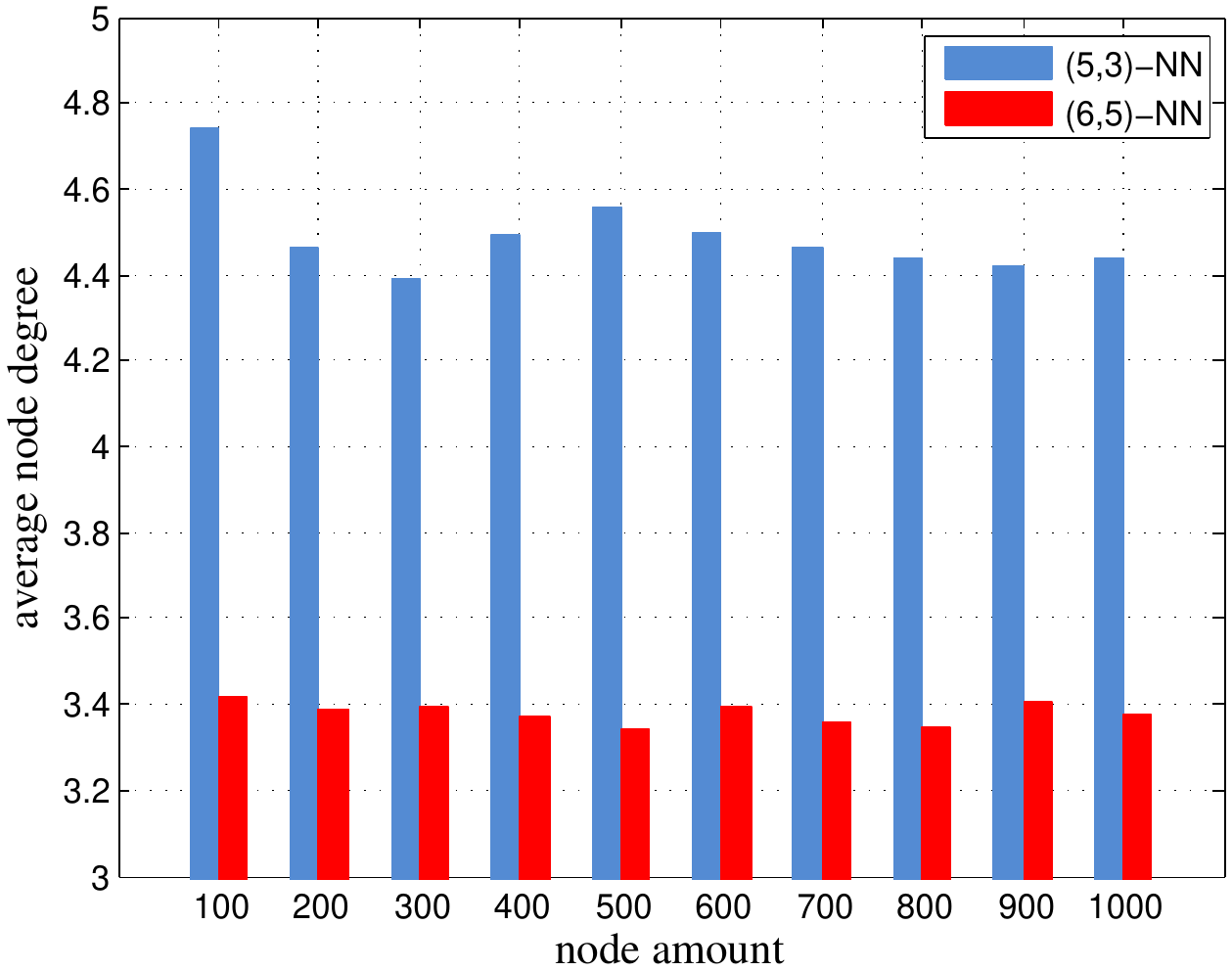}\\
  \caption{Average node degree: symmetric $(5,3)$-NN, $(6,5)$-NN graph}\label{fig:deg-53-65}
\end{figure}


\par Therefore, $(5,3)$-NN and $(6,5)$-NN graphs  are chosen for WSNs. Comparison results in terms of connectivity probability and average node degree are presented in Figure~\ref{fig:prob-53-65} and Figure~\ref{fig:deg-53-65}. For a small-scale (i.e. node amount is less than 1000) network topology, Figure \ref{fig:prob-53-65} shows that, when node amount exceeds $500$, the connectivity probability of symmetric $(6,5)$-NN graph remarkably decreases. In contrast, symmetric $(5, 3)$-NN graph evidently features  stubborn stability on connectivity probability. However, as shown in Figure \ref{fig:deg-53-65}, symmetric $(6, 5)$-NN graph has smaller average node degree.  Thus, $(k = 5, \; j = 3)$ or $(k = 6, \; j = 5)$ should be flexibly chosen for a specific application. In the rest of this paper, we select $(5, 3)$-NN graph for further experiments.

\par Analysis above demonstrates that the first $k$ nearest neighbors of a node exert disparate influence on guaranteeing network connectivity. Longer link within the first $k$ nearest neighbors makes more contribution in terms of network construction. For example, as numerical results demonstrate, connections among the last three of the first five nearest nodes $(k = 5, j = 3)$ competently build up a connected network.

\par
\begin{figure}[htbp]
  \centering
  \includegraphics[scale=0.6]{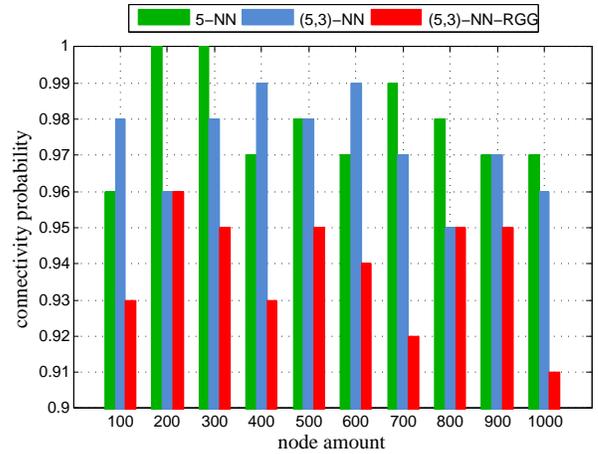}\\
  \caption{Connectivity probability: symmetric $5$-NN, $(5,3)$-NN, $(5,3)$-NN~-~RGG}\label{fig:c-prob-53s}
\end{figure}

\begin{figure}[htbp]
  \centering
  \includegraphics[scale=0.6]{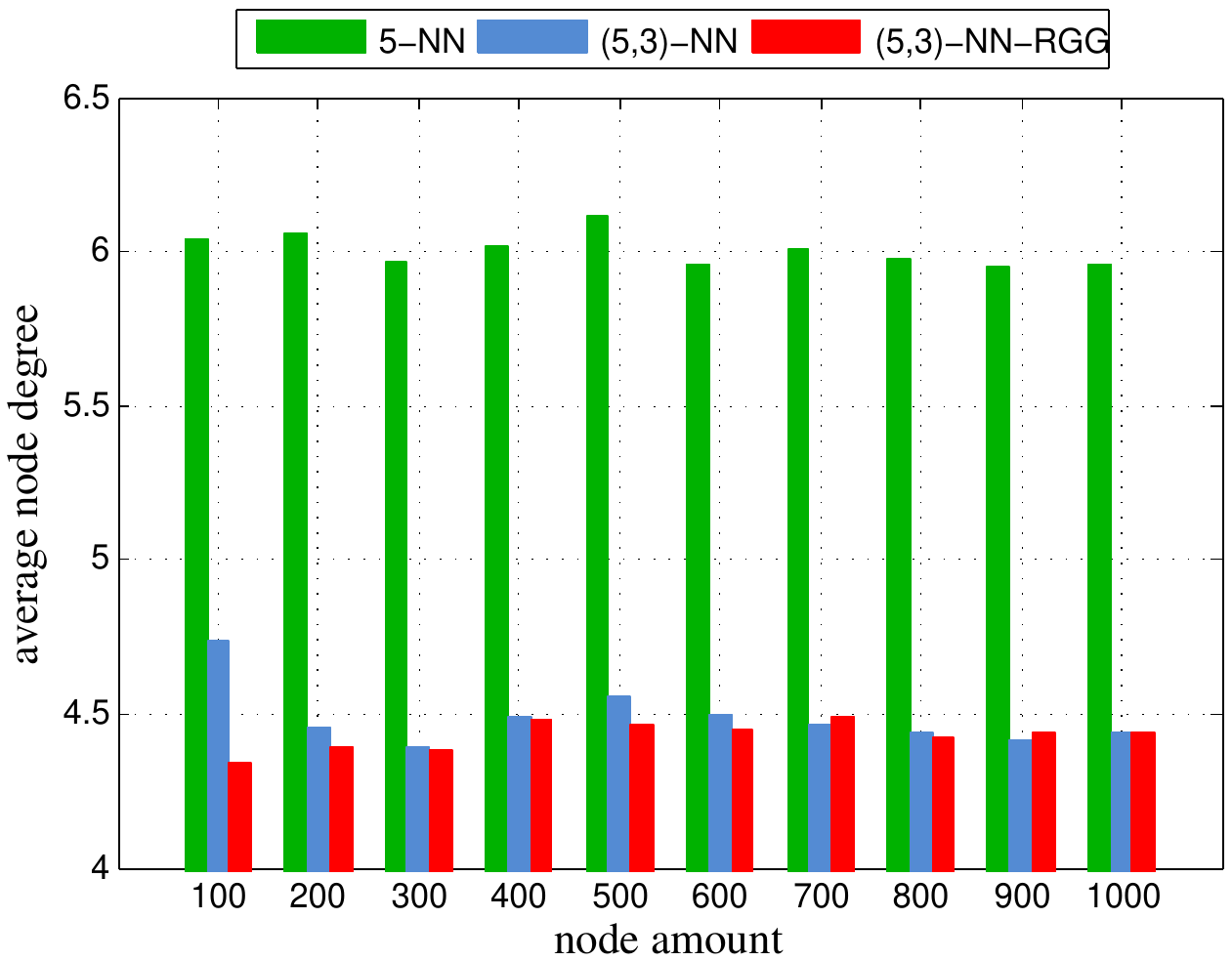}\\
  \caption{Average node degree: symmetric $5$-NN, $(5,3)$-NN, $(5,3)$-NN~-~RGG}\label{fig:c-deg-53s}
\end{figure}

 Last, we construct the composite topology combining symmetric $(k,j)$-NN and RGG. The constrained transmission radius $r_n$ for node amount $n$ is calculated by
\begin{eqnarray*}\label{radius}
  r_n &=& \sqrt{\frac{\log n + (2k-1)\log \log n + \xi}{\pi n}} \\
  \xi &=& \left\{
        \begin{array}{ll}
            -2\log \Big( \sqrt{e^{-\varsigma} + \frac{\pi}{4}}-\frac{\sqrt{\pi}}{2} \Big),   &   if \; k=1 \\
             2\log \frac{\sqrt{\pi}}{2^{k-1}k!} + 2\varsigma,                        &   if \; k>1
        \end{array} \right.
\end{eqnarray*}
where $k$ represents node degree, and $\varsigma$ is a constant \cite{wan2010asymptotic}. Here, $k = 5$, $\varsigma = 3$. Connectivity probability as well as average node degree of the composite topology is illustrated in Figure \ref{fig:c-prob-53s} and Figure \ref{fig:c-deg-53s}.

\par Figure \ref{fig:c-prob-53s} indicates that the connectivity probability of the composite topology, in comparison to that of symmetric $5$-NN as well as $(5,3)$-NN graph, decreases slightly to around $93.9\%$, which is still acceptable in practical applications. In Figure \ref{fig:c-deg-53s}, symmetric $(5,3)$-NN and the composite topology, comparing to symmetric $5$-NN graph, boast smaller average node degree, which may lead to less message collision probability and larger system throughput. The average node degree of the composite topology is $4.4316$. So we achieve the link gain about $(6.0063-4.4316) \times \frac{n}{2} = 0.78735n$, slightly larger than $0.75805n$ in symmetric $(5,3)$-NN graph.

\par
\begin{figure}[htbp]
\centering
\includegraphics[scale=0.6]{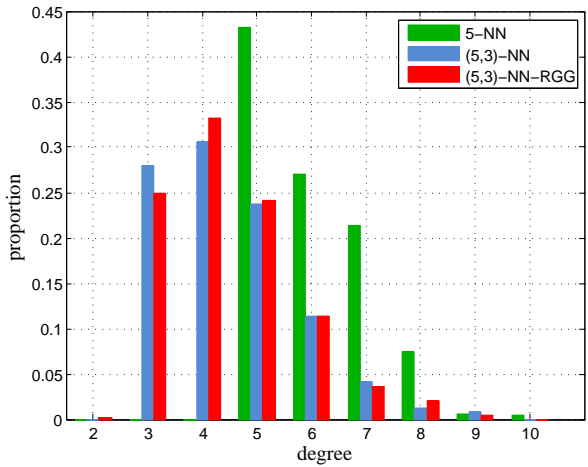}
\caption{Degree distribution (node amount $500$)}\label{fig:degree_distribution}
\end{figure}

 Further, Figure \ref{fig:degree_distribution} reveals the degree distribution among symmetric $5$-NN, $(5, 3)$-NN and the composite topology.\footnote{Figure \ref{fig:degree_distribution} describes a $500$ nodes scenario. For other node amounts, similar figures could be drawn.} We record the proportion of the number of the nodes with the same degree to node amount. Evidently, low degree nodes emerge more frequently in symmetric $(5,3)$-NN and the composite topology.

\par
\begin{figure}[htbp]
  \centering
  \includegraphics[scale=0.5]{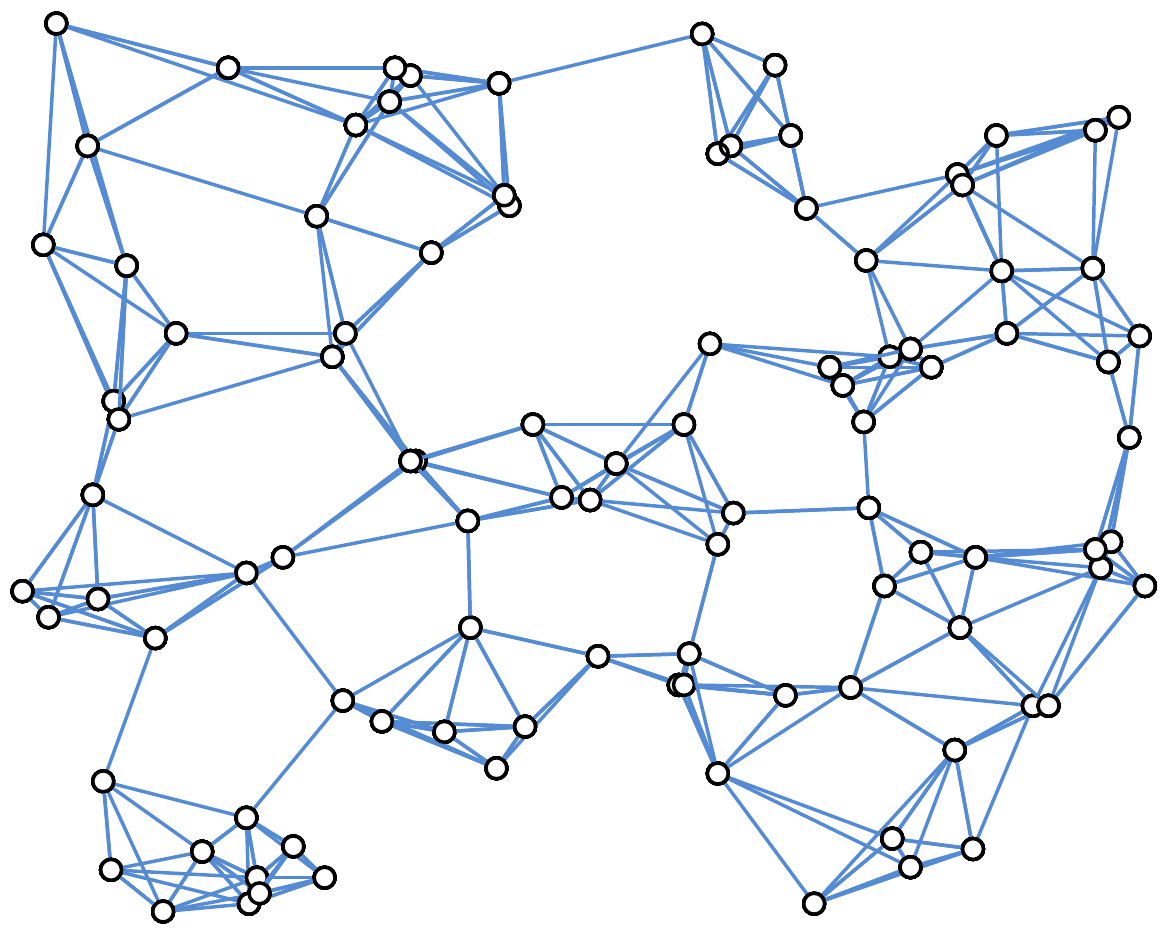}\\
  \caption{Symmetric $5$-NN graph ($100$ nodes)}\label{fig:sym-51-100}
\end{figure}

\begin{figure}[htbp]
  \centering
  \includegraphics[scale=0.5]{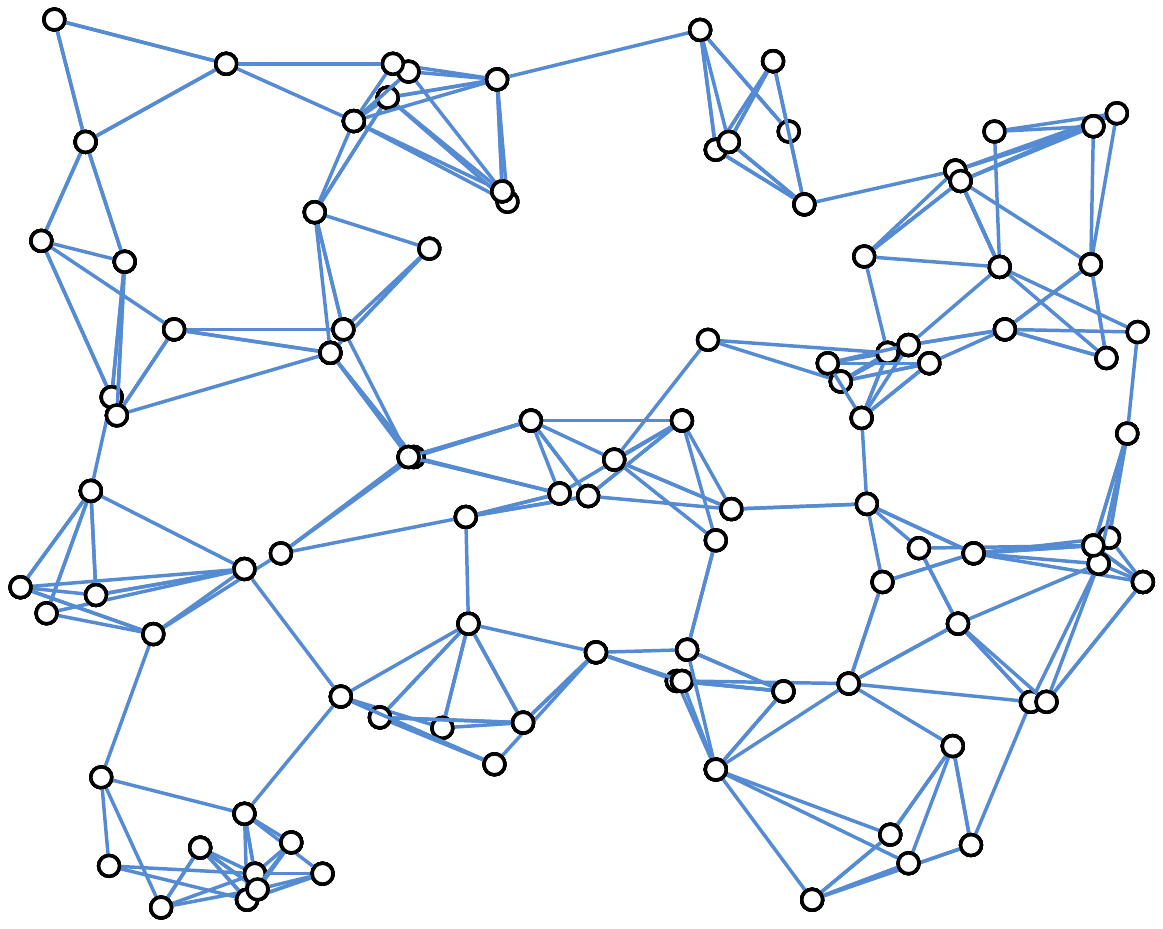}\\
  \caption{Composite topology ($(5,3)$-NN-RGG, $100$ nodes)}\label{fig:sym-53RGG-100}
\end{figure}

 In order to make a visual comparison between symmetric $5$-NN and our composite topology, we uniformly distribute $100$ nodes in a unit-area square, plotting Figure \ref{fig:sym-51-100} and Figure \ref{fig:sym-53RGG-100}. Notably, the composite topology (see Figure \ref{fig:sym-53RGG-100}), while generating a connected network topology, simplifies symmetric $5$-NN graph (see Figure \ref{fig:sym-51-100}) to a certain extent.

\section{Conclusion}
\par This paper has proposed several novel topologies for wireless sensor networks, which feature that less links can guarantee the network connectivity. For current all-neighbors-connected topologies, our results still have significant meaning, that is, the interruption of the shortest and/or second shortest links at several (even all) nodes, may not break off the connectivity.  However, a smarter route may be needed when such interruption happens, which is our future work.


%

\section*{Acknowledgments}
\par The authors acknowledge the financial support by the NSF of China under Grants 61103018 and 61472343, and the NSF of Jiangsu Province of China under Grant BK2012683.\@ J. Ding also acknowledges the support by SRF for ROCS, SEM, and Talent Project of New Century, Yangzhou University.

\ifCLASSOPTIONcaptionsoff
  \newpage
\fi



\bibliographystyle{IEEEtran}
%




\bibliography{reference_kjNN_RGG}

%








\end{document}